# Topology Control and Routing in Mobile Ad Hoc Networks with Cognitive Radios

Quansheng Guan*†, F. Richard Yu* and Shengming Jiang†
*Department of Systems and Computer Engineering, Carleton University, Ottawa, ON, Canada
†School of Electronic and Information Engineering, South China University of Technology, P.R. China

*Abstract*—Cognitive radio (CR) technology will have significant impacts on upper layer performance in mobile ad hoc networks (MANETs). In this paper, we study topology control and routing in CR-MANETs. We propose a distributed Prediction-based Cognitive Topology Control (PCTC) scheme to provision cognition capability to routing in CR-MANETs. PCTC is a midware-like cross-layer module residing between CR module and routing. The proposed PCTC scheme uses cognitive link availability prediction, which is aware of the interference to primary users, to predict the available duration of links in CR-MANETs. Based on the link prediction, PCTC constructs an efficient and reliable topology, which is aimed at mitigating re-routing frequency and improving end-to-end network performance such as throughput and delay. Simulation results are presented to show the effectiveness of the proposed scheme.

## I. INTRODUCTION

Cognitive radio (CR) [1] is an enabling technology to allow cognitive users (i.e., unlicensed users or secondary users) to operate on the vacant parts of the spectrum allocated to licensed users (i.e., primary users). CR is widely considered as a promising technology to deal with the spectrum shortage problem caused by the current inflexible spectrum allocation policy. It is capable of sensing its radio environment, and adaptively choosing transmission parameters according to sensing outcomes, which improves cognitive radio system performance and avoid interfering primary users [2]–[4].

Recent research activities conducted in CR are mainly focusing on the opportunistic spectrum access and physical layer transmission throughput [3]. However, CR technology will have significant impacts on upper layer performance in wireless networks, especially in mobile ad hoc networks (MANETs), which enable wireless devices to dynamically establish networks without necessarily using a fixed infrastructure. Certainly, issues in non-cognitive MANETs in general are still of interest in the CR paradigm. However, some distinct characteristics of CRs introduce new non-trivial issues to CR-MANETs [5]–[8].

One of the particularly important networking issues is routing in CR-MANETs. From routing perspective, it is expected that data packets are routed via a stable and reliable path to avoid frequent re-routing problem, since frequent re-routing may induce broadcast storm to the network, waste the scarce radio resources and degrade end-to-end network performance such as throughput and delay [9]–[13]. Compared to classical MANETs, a path in CR-MANETs is especially unstable, since it is not only affected by node mobility among cognitive users but also by the interference to primary users. We argue that routing in CR-MANETs (we call it cognitive routing hereafter) should have the following unique characteristics:

- Primary user interference awareness: Cognitive routing should choose a path that the interference to primary users is below the required threshold.
- Link availability prediction: It is pointed out in [14] that a CR network should be forward-looking, rather than reactive. Indeed, since spectrum sensing may take a long time and be delayed [3], a reactive CR network will degrade the performance. Therefore, cognitive routing should not only be aware of primary users, but also foresee link available period in terms of its interference to primary users.
- Adaptive acting: Cognitive routing should be adaptive to choose a path based on the prediction to mitigate re-routing frequency, and then increase end-to-end throughput and decrease end-to-end delay.

There are a large number of routing protocols proposed in classical MANETs, such as DSDV [15], DSR [16], AODV [17], and etc. However, it is difficult to apply them directly to CR-MANETs due to the distinct characteristics described above. On the other hand, it may not be desirable to design a new routing protocol dedicated to CR-MANETs due to the maturity and availability (e.g., source code for different operating systems) of existing routing protocols. From this point of view, it is better to provision the cognition capability to routing via a middle-ware-like mechanism.

In this paper, we use topology control to achieve this objective. Topology control is originally developed for wireless sensor networks [18], MANETs [19]–[21] and wireless mesh networks [22] to reduce energy consumption and interference. It works as a middle ware, connecting routing and lower layers. Topology control focuses on network connectivity with the link information provided by medium access control (MAC) [23], [24] and physical layers, which both belong to cognitive radio module in CR-MANETs. When constructing network topology, topology control takes care about the interference to primary users and link availability prediction in CR-MANETs. Specifically, we propose a novel scheme, called Prediction-based Cognitive Topology Control (PCTC), to provision cog-nition capability to routing protocols in CR-MANETs in a distributed manner. Two distinct features of the proposed scheme are as follows:

1) It can predict both link duration and the probability that this duration may really last to the end of this period, considering both user mobility and interference to primary users.
2) The proposed topology control scheme is performed in a distributed manner. Since it is not practical to collect global information in CR-MANETs, a distributed topology control scheme is desirable. In other words, network connectivity is preserved in a distributed manner.

Simulation results indicate that the proposed PCTC scheme results in simpler topology and has longer link duration than that without topology control. Routing protocols also performs better in the resulting topology.

The rest of the paper is structured as follows. Section II presents the topology control and routing problems in CR-MANETs. A cognitive link availability prediction scheme is presented in Section III. With this link prediction scheme, Section IV presents the distributed topology control and routing schemes for CR-MANETs. Simulation results are presented and discussed in Section V. Finally, Section VI concludes this study.

## II. TOPOLOGY CONTROL AND ROUTING IN COGNITIVE RADIO MOBILE AD HOC NETWORKS

The dynamic spectrum availability and the importance of limiting the interference to primary users (PUs) differentiate CR-MANET from classical MANETs. In CR-MANETs, two factors affect spectrum availability. The first one is primary user activity. Since cognitive users (CUs) are considered as low priority and they are secondary users of the spectrum allocation to primary users, cognitive users should sense the spectrum to detect primary user activity. Due to the capability limitation of some cognitive users and a large number of possible spectrum bands, spectrum sensing may take a long time and be delayed [3]. The second factor that affects spectrum availability is cognitive user mobility. Spectrum becomes unavailable when a node moves into the interference boundary of an active PU. Therefore, correctly inferring mobility conditions in designing effective routing and topology control schemes is very important in CR-MANETs [5].

In this sense, we concentrate on how the spectrum availability due to CU node mobility affects the network performance in CR-MANETs. We use a simple scenario shown in Fig. 1 to illustrate the effects of CU node mobility on routing in CR-MANETs. There are one PU and four CUs in this scenario. The source communication node is CU1, and CU4 is the destination node. The arrows indicate the movement directions of CU2 and CU3. As we can see from Fig. 1, CU3 is moving towards the PU, while CU2 is moving away from the PU. In this scenario, if we use existing routing protocols developed in classical MANETs, such as DSDV, DSR, AODV, a shortest path criteria will be used to select paths. Link CU1→CU3 may be selected in the route. However, as CU3 moves close to the PU, Link CU1→CU3 cannot be used due to the interference from CU3 to the PU. The procedure of re-launching route request will be needed to find another route. By contrast, Link CU1→CU2 does not have this problem, since CU2 is moving away from the PU and the interference from CU2 to the PU is below the threshold. The re-routing problem would not happen if Link CU1→CU2 instead of CU1→CU3 was selected. From this example, we can see that mobility prediction can help improve the performance of cognitive routing in CR-MANETs.

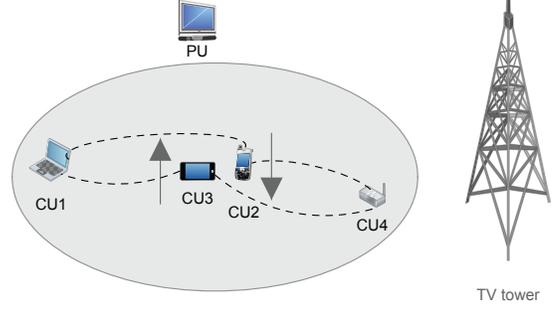

Fig. 1. A motivation scenario: cognitive routing selects CU2 in its path other than CU3 for the sake of reducing re-routing frequency when cosidering interference avoidance to PU. The arrows indicate the movement directions of CU2 and CU3.

It is desirable that cognitive routing should favor links between CUs with long duration to prolong the path survival time and improve the path stability. Existing routing protocols in classical MANETs, such as DSDV [15], DSR [16], AODV [17], lack this cognitive capability. Nevertheless, these routing protocols are relatively mature, and the source code is widely available in different operating systems. Therefore, to provision cognition capability to routing, we adopt a midware-like technology, topology control, in CR-MANETs. In our framework, topology control is treated as a cross-layer module connecting routing layer and CR module. Since routing is conducted on a given topology, topology control is essential to provision cognition capability to routing protocols. An intrinsic feature, also a key challenge of cognitive topology control, is to be intelligent enough to make routing protocols forward-looking, which is also one of the requirements of CR-MANETs. In our scheme, topology control can construct a reliable topology based on PU-aware link availability prediction, which is described in the following section.

## III. COGNITIVE LINK AVAILABILITY PREDICTION

A reasonable link availability prediction model for classical MANETs has been proposed in our previous work [25]. The basic principle of the prediction framework is to provide a predicted time period $T_p$ that the link between two nodes will stay available. In addition, another important parameter, denoted by $L(T_p)$, can be obtained from this framework to estimate the probability that this link may really last to the end of $T_p$.

In CR-MANETs, only $[T_p, L(T_p)]$ is not sufficient to predict link duratin, since the links between CUs are affected not only by CU mobility but also by the interference to PUs. In this situation, a link with a high quality may be considered unavailable, if a node in that link is moving close to a primary user. As a result, in order to avoid interference, the distance between a cognitive user and primary user should be monitored. This paper proposes another pair of $[\hat{T}_p, L(\hat{T}_p)]$ to predict the link availability before nodes are moving into the interference boundary of primary users. Similar to $T_p$, $\hat{T}_p$ is the predicted time till a CU moves into the interference area of a PU, and $L(\hat{T}_p)]$ is its corresponding probability. Then the final link availability is revealed by the combination of $[T_p, L(T_p)]$ and $[\hat{T}_p, L(\hat{T}_p)]$ together to enable cognitive link prediction.

### A. $[\hat{T}_p, L(\hat{T}_p)]$ Estimation

Under the assumption that velocities of nodes stay constant during each random epoch, it is easy to get that $\partial^2 d^2/\partial^2 T = 0$ [25], [26], where $d$ presents the link distance between two nodes and $T$ is the time interval. Then $d$ can be expressed as

$$d^2 = \alpha T^2 + \beta T + \gamma, \quad (1)$$

where $\alpha$, $\beta$ and $\gamma$ are constant and they can be obtained by three points of measurements $(t_0, d_0)$, $(t_1, d_1)$ and $(t_2, d_2)$. The sample time $t_i = t_0 + T_i$ and $d_i$ is the corresponding measured distance between two nodes.

For two nodes in each other's transmission range, as long as their velocities remain constant and are not the same as each other, they will be sure to travel out of this range. In another words, there exists a solution $T_p$ for (1). Different from $T_p$ prediction, some nodes may be moving far away from primary users, and they will always stay outside of the interference boundary. In this case, there exist no solution $\hat{T}_p$ for (1). If a cognitive user is in the interference region of a primary user, which is detected by CR module, it is not necessary to prediction its corresponding link availability. $\hat{T}_p$ is set to 0 in this case. Therefore, we only discuss the situation where nodes are out of interference boundary initially. Let $\rho$ denote the interference boundary radius, then $\Delta = \beta^2 + 4\alpha\rho^2 - 4\alpha\gamma$. To make sure there exists at least a solution $\hat{T}_p$ for (1), we have $\Delta \geq 0$. In addition, note that $\hat{T}_p \geq 0$. According to [25], the maximum allowable period $\hat{T}_p$, if counted from $t_2$, is calculated by

$$\hat{T}_p = \begin{cases} \frac{\sqrt{\beta^2 + 4\alpha\rho^2 - 4\alpha\gamma} - \beta}{2\alpha} - t_2 & , \text{if} \Delta \geq 0 \text{ and } \Delta - \beta^2 \geq 0 \\ \infty & , \text{otherwise.} \end{cases} \quad (2)$$

Obviously, the preciseness of $\hat{T}_p$ depends on the measurement of $\rho$. There are some methods to obtain distance information. Typically, it may be derived from a path loss model and a measured received signal strength, or directly by Global Position Systems (GPS).

Similar to $[T_p, L(T_p)]$ estimation, we also have to provide a probability $L(\hat{T}_p)$ to $\hat{T}_p$. An enhanced method of $L(T_p)$ in [27] is adopted to derive $L(\hat{T}_p)$,

$$L(\hat{T}_p) \approx e^{-\lambda \hat{T}_p e^{-\lambda \tau}} + \zeta(1 - e^{-\lambda \hat{T}_p}), \quad (3)$$

where the parameters $\tau$ and $\zeta$ are obtained by measurement. While considering only the first change in velocities during period $T_p$ in [25], the equation (3) takes into account all possible changes in velocities that may happen during period $\hat{T}_p$. The detailed derivation of (3) can be found in [27]. Then, the pair of $[\hat{T}_p, L(\hat{T}_p)]$ is used to predict link duration corresponding to the interference to primary users.

### B. Link Availability Prediction

A link is considered available if the two nodes associated with this link are within the transmission range of each other and if they are both out of interference region of any primary user in a CR-MANET. The former is a common condition in MANETs while the latter is the specific requirement in CR-MANETs. Consequently, link availability should be determined by $[T_p, L(T_p)]$ and $[\hat{T}_p, L(\hat{T}_p)]$ together.

The authors of [25] and [28] use $T_p \times L(T_p)$ as a routing metric to assist routing protocols in selecting reliable paths. However, this is far from sufficient for cognitive MANETs. The available duration of a link in CR-MANETs, denoted by $T_a$, should be set to

$$T_a = \min_{i=1,2; j \in \{PUs\}} \{T_p \times L(T_p), \hat{T}_{p_i}^j \times L(\hat{T}_{p_i}^j)\}, \quad (4)$$

where $i$ is associated with the two ends of a link and $\{PUs\}$ is the set of PUs present in the network. The subscript $i$ and superscript $j$ indicate that a link will be unavailable if any of its ends moves into the interference region of any primary users.

## IV. COGNITIVE TOPOLOGY CONTROL AND ROUTING

To avert using the links with low duration such that they cannot avert frequent re-routing occurrence, in this section we will present the distributed topology control and routing schemes in CR-MANETs based on the link prediction presented above.

### A. Distributed Topology Construction

The dynamic changes due to CU mobility or the interference to PUs, which will result in frequent re-routing, waste large amount of scarce network resource, and achieve low end-to-end performance. The proposed PCTC algorithm aims to solve this problem by constructing a more reliable topology for routing protocols.

In our proposed cognitive topology control algorithm, a new link reliability metric is defined for topology construction. We first introduce a re-routing penalty denoted by $\delta$. This penalty is a time period which is incurred by re-routing and reduces link availability to $(T_a - \delta)$ in the sequel. Actually, the path duration is not the main factor that an end-to-end transmission concerns. Instead, in nature, it is expected to deliver as many packets as possible before a path failure happens. From this point of view, the only consideration of link available duration of $T_a$ is not enough since a link with long duration may have a poor quality resulting in low data rate. In the long run, it needs less re-routing if the paths selected consist of links with longer $T_a$ and higher quality. To quantify this measurement, we set the link weights according to

$$w = r \times (T_a - \delta), \tag{5}$$

where the link data rate $r$ captures its link quality. Herein, the re-routing penalty $\delta$ is converted into a capacity loss $r \times \delta$ during the available period. The link weight then presents the traffic carrying ability of this link. Based on (5), we define a path weight as

$$\mathcal{W} = \min w_i, i \in \mathbb{L}, \tag{6}$$

where $\mathbb{L}$ is the link set including all the links along the path. We can see that the definitions of link and path weights reflect not only the duration but its true data transmission ability. A link with long duration but bad link quality or a link with short duration though good quality will result in weak data transmission ability. It is desirable to transmit more data traffic before link failure.

The principle of PCTC is to preserve the reliable path with maximum path weight in (6) for any pair of nodes under connectivity guarantee. Similar to other topology control algorithms, the topology construction process in this paper consists of three steps: neighbor collection, path search and neighbor selection. The distributed Localized Dijkstra Topology Control (LDTC) algorithm in our previous work [29] to construct a energy-efficient topology has some beneficial properties, particularly the 1-spanner property, which preserves the global minimum energy paths in a distributed manner. In this paper, the distributed cognitive PCTC algorithm enhances LDTC to preserve the end-to-end reliable paths for CR-MANETs. As a distributed algorithm, each node runs the following procedure as an initial node:

1) Collect all of the neighbors, and calculate the edge weights according to (5).
2) Set the path weights to infinity for initial node and to zero for neighbors. Mark all the neighbors as unvisited and the initial node as the current one.
3) Calculate the path weights according to (6) from the initial node to unvisited neighbors via the current node. If they are larger than the previously recorded ones, update the path weights to them.
4) Mark the current node as visited and set the unvisited node with largest path weight as the current node. Then repeat from 3) until all the neighbors are visited.
5) All of the most reliable paths from the initial node are now found. The resulting topology is the output by preserving all the one hop neighbors of the initial node along these paths.

From the PCTC running procedures, we know that the reliable links are preserved and links with low duration may be removed in the resulting topology, resulting in more reliable and stable topology. In the follow subsection, we will study the properties of PCTC.

### B. Properties of the Proposed Topology Control Scheme

The cognitive topology control in this paper circumvents using the links with small availability and poor link quality under the condition that there exist better paths to replace these links. Accordingly, network connectivity is guaranteed. More importantly, all the global reliable paths are preserved. The computational complexity of the algorithm is $O(n \log n)$, which is a polynomial time. PCTC has the following interesting properties.

*Property 1 (Connectivity):* PCTC preserves the global network connectivity.

*Property 2 (1-Spanner):* The PCTC algorithm preserves global reliable path between any two nodes, that is, it has a spanner factor of 1 regarding the path weight defined in (6). A reliable path is referred to a path with the maximum path weight according to (6) and spanner factor is defined as

$$s(u,v) = \frac{\text{Reliable path weight in resulting topology}}{\text{Reliable path weight in original topology}}. \tag{7}$$

*Proof:* Suppose $path(u,v) = \{u_0 = u, u_1 = u_1, \cdots, u_k = v\}$ is the most reliable path connecting node $u$ and $v$, then any subpath $\{u_i, \cdots, u_j\}$, for $0 \leq i < j \leq k$, is also a reliable path. Therefore, if PCTC preserves all the local reliable paths, global reliable paths are sure to be preserved. Under the definition of (7), PCTC gets a 1-spanner factor. $\square$

According to the estimation framework in [25] and the cognitive link availability prediction in (4), we know that each node associated with a link has the same prediction on $T_a$. In addition, if they also have the same link data rate $r$, which is determined by link quality estimation, they get a symmetric link weight in the light of (5). As a result, paths are symmetric.

*Property 3 (Symmetry):* In the resulting topology by PCTC, a node $u$ is a neighbor of node $v$ while the node $v$ is also a neighbor of node $u$.

*Proof:* Given two adjacent nodes $u$ and $v$, and their neighbor sets $N_u$ and $N_v$ respectively, suppose $v \in N_u$ and $u \notin N_v$, there doesn't exist a path with higher weight from $u$ to $v$ than that of Link $u \to v$. On the other hand, from node $v$'s point of view, there exists another reliable path to $u$ other than Link $v \to u$. Consequently, $v$ is excluded from the neighborhood of node $u$ by PCTC owning to link symmetry. This contracts the assumption. $\square$

What we concern is to what extent the topology control algorithm controls the topology. Let $\phi$ stand for the node degree in the resulting topology. Then

$$\phi = \sum_{i=1}^{n} 1_{\{e_i\}}, \tag{8}$$

where $e_i$ is used to present the event that a new neighbor is added to the final topology in the *ith* iteration and $n$ is the original node degree. If we define the control intensity of a topology control algorithm as the mean ratio between the resulting node degree and the original node degree, we get the following property.

*Property 4 (Control intensity):* The control intensity of PCTC reaches

$$\mathbb{E}[\frac{\phi}{n}] = \frac{\sum_{i=1}^{n} \mathbb{P}\{e_i\}}{n}. \quad (9)$$

In each iteration, PCTC calculates the weights of extended paths from the current node to the unvisited nodes and then updates the reliable path weights from the initial node to the unvisited, namely, $\vec{l}$ vector. Once a node is marked as visited, its reliable path to the initial node will not be changed in the sequel, that is, there does not exist another path that has larger weight in the following iterations. In each iteration, the node with the largest path weight in $\vec{l}$ vector is marked as visited and acts as the current node for the next iteration, i.e.,

$$k^* = \arg \max_{k \in unvisited} \vec{l}(k). \quad (10)$$

Only one unvisited node satisfies the above equation is changed to visited and the corresponding link is added to resulting topology for each iteration. In fact, the added link has the largest weight among the links between visited nodes and unvisited nodes Hence, whether $e_i$ happens or not depends on the other end of the added link. If it is the initial node, $e_i$ happens, and vice versa. Obviously, $e_i$ for $i = 1, 2, \cdots, n$, is independent of each other.

In the *i*th iteration, there have been $i$ visited nodes and $(n - i)$ unvisited nodes. One end of he added link in this iteration is an unvisited node and the other is a visited node. As aforementioned, $e_i$ will happen if the visited node is the initial node. The $e_i$ probability is then given by $\mathbb{P}\{e_i\} = \frac{1}{i}$. Accordingly, we obtain

$$\mathbb{E}[\frac{\phi}{n}] = \frac{1}{n} \cdot \sum_{i=1}^{n} \frac{1}{i}. \quad (11)$$

*C. Cognitive Routing on the Resulting Topology*

Routing is defined to select paths in a network to send data traffic. A challenge in CR-MANETs is to provision cognition capability to routing protocols. With the proposed PCTC, existing routing protocols can be easily adopted in CR-MANETs with the cognitive capability. Based on the cognitive link prediction scheme presented in Section III, routing on the PCTC resulting topology is aware of primary users and forward-looking to link duration. Further, the topology control scheme makes routing adaptive to mobile environment, in which cognitive routing favors reliable paths in the network. Take the popular routing protocols, such as DSR and AODV, for examples. They usually send routing request packets, called RREQs, to find a path for a source and a destination. When a RREQ reaches at an intermediate node, it may be dropped if the transmitter doesn't exist in the neighborhood relationship generated by the PCTC algorithm. Otherwise, this RREQ is disposed by the intermediate node and re-broadcasted again. As a result, the links in the vicinity of primary users or with poor quality and short available duration are avoided. The first RREQ is replied in terms of a routing metric of first found path to confirm the found path.

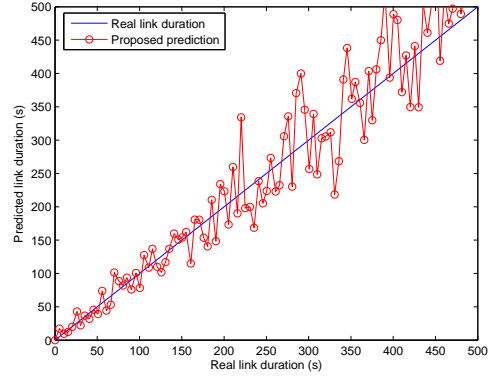

Fig. 2. Predicted link duration v.s. real link duration in the proposed cognitive link prediction scheme.

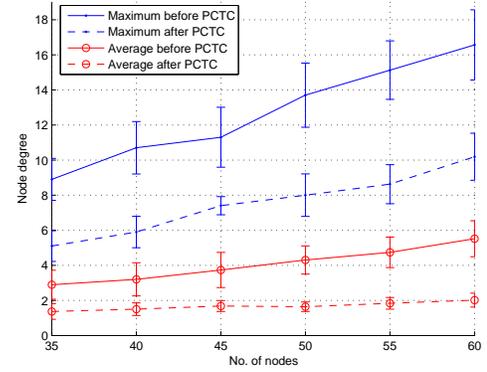

(a) Node degree

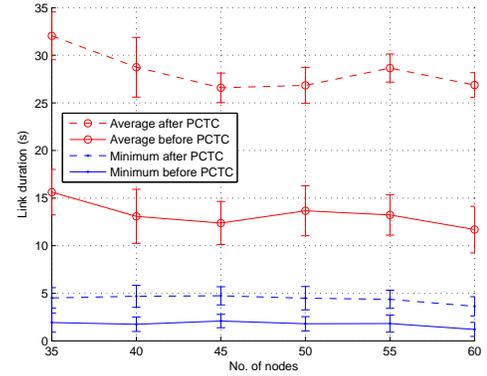

(b) Link duration

Fig. 3. Properties of PCTC resulting topologies. Confidential level is 95%.

Actually, PCTC changes neighborhood relationship among nodes and assigns each path a weight with regard to reliability. Based on the proposed cognitive topology control scheme, we can improve the performance of other routing metrics such as shortest path or QoS routings. The performance of routing on the resulting topology is studied by computer simulations in Section V.

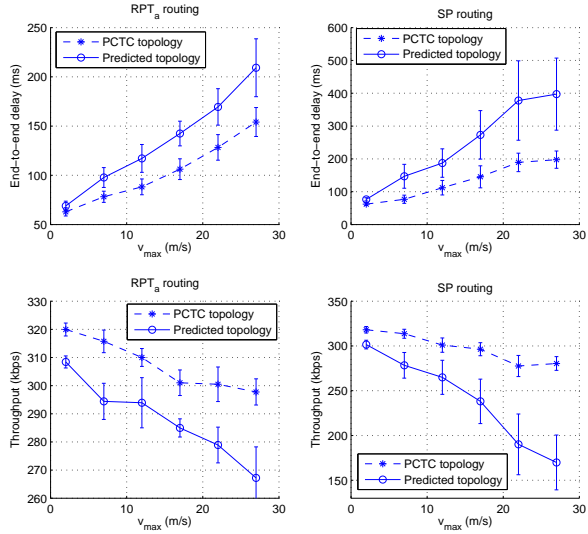

Fig. 4. End-to-end performance studies over different maximum velocities in mobility model. Confidential level is 95%.

## V. SIMULATION RESULTS AND DISCUSSIONS

In this section, the proposed scheme is evaluated via computer simulations. In the simulation environment, nodes are randomly moving in a 2-dimensional space according to the random-walk-based mobility model, where a node moves with a direction uniformly and a speed uniformly from 0 to $v_{max}$ with exponentially distributed epochs. IEEE 802.11 is adopted for the MAC layer. The maximum transmission range of each node is 300 meter, and the transmission rate is 2 Mb/s.

We first look into the accuracy of the prediction-base cognitive link duration prediction. Fig. 2 shows the prediction results for space $500 \times 500~m^2$ and mean epoch $\lambda^{-1} = 60s$. It is revealed that $T_a$ fluctuates around $T_r$, which is the real time period that a link is continuously available corresponding to the predicted $T_a$.

With the predicted link duration, we further evaluate the performance of the prediction-base cognitive topology control algorithm.

Fig. 3 shows detailed properties about the node degree and link duration of the topology control algorithm. We run the simulations for 100 times. The confidence level is 95%. From Fig. 3(a), we can see that the resulting topology has smaller average node degree and maximum node degree than those without topology control. Small node degree will mitigate contention in the shared wireless medium and the number of RREQs in the route discovery process. We note that average node degree retains a small value as number of nodes in the network increases, which also makes network scalable. In addition to node degree, the topology control algorithm also results in longer link duration, which is shown in Fig. 3(b). This indicates that the resulting topology is more stable and it is possible to reduce re-routings in the network.

It has been proven that the resulting topology is provisioned with cognition capability to avoid interference to primary users. Routing over the resulting topology is then adaptive in CR-MANETs. The following two routing metrics are adopted to evaluate the performance of the topology control algorithm further in the simulations.

- Shortest Path (SP): Each flow is transmitted along the path with the minimum number of hops.
- Reliable Path with regard to $T_a$ ($RPT_a$): This is the proposed scheme in this paper. It selects the path with maximum path duration according to (6).

End-to-end performance is a critical objective of routing protocols. Fig. 4 demonstrates the two main performance metrics, end-to-end throughput and delay, of routing on the predicted topology and cognitive routing on the PCTC resulting topology. For $RPT_a$ routing, though the same path is selected on the predicted topology and PCTC topology due to Property 2, end-to-end throughput and delay of cognitive routing are still better than those without topology control. This is because less node degree in PCTC topology may alleviate the contention in shared wireless medium and results in less medium access time. As for SP routing, though paths with more hop counts, the end-to-end throughput under the resulting topology is higher than that under $T_a$ topology. The end-to-end delay is also shorter. The reason exists in that the paths under the resulting topology are with longer duration, resulting in less link breakages and the less re-routing frequency.

## VI. CONCLUSIONS AND FUTURE WORK

Cognitive radio technology will have significant impact on upper layer performance in cognitive radio mobile ad hoc networks (CR-MANETs). In this paper, we have proposed a framework to provision cognition capability to the routing protocols in CR-MANETs, which are aware of the interference to primary users, forward-looking to link available duration and adaptive to mobility environment. A novel distributed prediction-based cognitive topology control scheme was presented. Link available duration prediction takes into account both the interference to primary users and the mobility of cognitive users. We have evaluated the routing performance over the resulting topology by computer simulations. It was shown that, with the proposed mid-ware-like topology control residing between the routing layer and CR module, the resulting network topology is simpler and links have longer durations. Classical MANET routing protocols can be easily adopted in CR-MANETs with the cognitive capability.

Future work is in progress to consider the effects of topology control and routing on multimedia applications in cognitive radio networks. We also plan to consider security issues [30] and admission control issues [31], [32] in heterogeneous networks [33], [34].